\pdfoutput=1
\documentclass[
showkeys,useAMS,twocolumn,nofootinbib,pra]{revtex4-1}
\usepackage{graphics, graphicx, amsmath, amssymb}
\usepackage{epstopdf}
\usepackage{dsfont}
\usepackage{hyperref}
\usepackage{color}
\usepackage[normalem]{ulem}
\usepackage[T1]{fontenc}
\usepackage[utf8]{inputenc}

\newcommand{\eq}[1]{\begin{equation}\begin{aligned}#1\end{aligned}\end{equation}}
\newcommand{\iu}{\text{i}}
\newcommand{\eu}{\text{e}}
\newcommand{\ha}{\hat{a}}
\newcommand{\had}{\hat{a}^\dagger}

\newcommand{\haL}{\hat{a}_L}
\newcommand{\haLd}{\hat{a}^\dagger_L}
\newcommand{\haR}{\hat{a}_R}
\newcommand{\haRd}{\hat{a}^\dagger_R}

\newcommand{\ket}[1]{\left|#1\right\rangle}
\newcommand{\bra}[1]{\left\langle#1\right|}

\newcommand{\vac}{\ket{\text{vac}}}
\newcommand{\id}{\hat{\mathds{1}}}
\newcommand{\ot}{\otimes}
\newcommand{\del}[1]{}
\newcommand{\add}[1]{\color{black}{#1}\color{black}}

\begin{document}
\title{Quantum theory of polarimetry: From quantum operations to Mueller matrices}
\date{\today}
\author{Aaron Z. Goldberg}
\email{goldberg@physics.utoronto.ca}
\affiliation{Department of Physics, University of Toronto, Toronto, ON, M5S 1A7}
\begin{abstract}
Quantum descriptions of polarization show the rich degrees of freedom underlying classical light. While changes in polarization of light are well-described classically, a full quantum description of polarimetry, which characterizes substances by their effects on incident light's polarization, is lacking.
We provide sets of quantum channels that underlie classical polarimetry and thus correspond to arbitrary Mueller matrices. This allows us to inspect how the quantum properties of light change in a classical polarimetry experiment, and to investigate the extra flexibility that quantum states have during such transformations. Moreover, our quantum channels imply a new method for discriminating between depolarizing and nondepolarizing Mueller matrices, which has been the subject of much research. This theory can now be taken advantage of to \add{provide quantum enhancements in }\del{improve }estimation strategies \add{for }\del{in }classical polarimetry and to further explore the boundaries between classical and quantum polarization.
\end{abstract}
\maketitle


Polarization of light has been studied for centuries \cite{Bartholin1670,Stokes1851,Chandrasekhar1950,Goldstein2003,GilOssikovski2016}\add{\cite{AzzamBashara1977}}. Some materials transmit light of one polarization better than another, others change the polarization of light as it passes through \cite{Collett1992}, and these qualities are sufficient to discriminate important substances in fields ranging from oceanography \cite{VossFry1984} to biology \cite{Boulvertetal2004,Ghoshetal2009,Tuchin2016} to astronomy \cite{Dulketal1994,Tinbergen2005}. 
Known as polarimetry, characterizing transformations of incident light's polarization as a proxy for characterizing materials has found far-reaching applications in classical optics \cite{BornWolf1999}. We here investigate the nuances added to polarimetry by a quantum description of light. 

Polarization has many applications in quantum optics. Each individual photon, analogous to a classical state of light \cite{FalkoffMacDonald1951,Aielloetal2007,Galazoetal2018}, has its own polarization \cite{Beth1935,Allenetal1992}. It is further possible for quantum states to appear ``classically unpolarized'' yet possess ``hidden'' polarization properties, leading to novel quantum-mechanical polarization effects \cite{PrakashChandra1971,Klyshko1992,Klyshko1997,Tsegayeetal2000,Usachevetal2001,Bushevetal2001,Luis2002,AgarwalChaturvedi2003,Bjorketal2015,LuisDonoso2016,ShabbirBjork2016,GoldbergJames2017,Bouchardetal2017,GoldbergJames2018Euler}.  
 The photon's polarization degree of freedom is highly useful in quantum information and quantum communication \cite{NielsenChuang2000}, having been used for applications such as enhancing measurement sensitivities \cite{Mitchelletal2004} and distributing quantum keys \cite{Gisinetal2002,Tangetal2014}. The applications of quantum polarization are constantly expanding.

One pertinent application is the ability to use the quantum polarization properties of light to increase measurement resolution \add{relative to the classical shot-noise limit \cite{Caves1981} } in specific kinds of polarimetry \cite{GoldbergJames2018Euler}. Using carefully designed quantum states of light that seem unpolarized from a classical point of view, one can dramatically enhance the sensitivity of rotation measurements, which constitute a subset of polarimetric measurements \cite{GoldbergJames2018Euler}. \add{Such a measurement could be used to detect fragile compounds such as biological samples \cite{AzzamBashara1977} whose optical activity has previously been masked by nonlinear effects at high intensities and by shot noise at low intensities. } \add{The quantum advantage for rotation measurements is achieved in the quantum Fisher information paradigm \cite{Helstrom1967,Helstrom1973,Holevo1973,Holevo1976}, in which the best \textit{possible} measurement sensitivity can be found without having to specify a specific measurement apparatus \cite{Giovannettietal2004,Nagaoka2005,Paris2009,TothApellaniz2014,DemkowiczDobrzanskietal2015,Szczykulskaetal2016,SidhuKok2019,Liuetal2019,Albarellietal2020}; the quantum Fisher information depends only on the input states for a given polarization transformation. }
However, a full quantum-mechanical description of polarimetry does not yet exist beyond single- and two-photon models \cite{Aielloetal2007}\add{, which have been probed by recent experiments \cite{Altepeteretal2011,Yoonetal2020}}. We here provide a description of polarimetry using quantum-mechanical operators acting on quantum states of light, in order to facilitate a deeper understanding of polarization of light and its applications. This is important for finding new quantum enhancements in \add{shot-noise-limited } polarimetry as well as for discovering the constraints placed by quantum mechanics on classical polarimetry.

To do so, we start by giving a classical explanation of polarization. Polarization is described by the four Stokes parameters, and polarimetry characterizes the $4\times4$ Mueller matrix governing their linear transformations, constraints on which are well studied \cite{FryKattawar1981,GilBernabeu1985,Cloude1986,Barakat1987,Cloude1990,GivensKostinski1993,Nagirner1993,vanderMee1993,Raoetal1998,Gil2000,vanZyletal2011,Giletal2013,Zakerietal2013}. These classical Stokes parameters then correspond to expectation values of non-commuting quantum operators \cite{Fano1949,Fano1954,JauchRohrlich1955}, leading to a rich higher-order polarization structure. We describe the quantum channels that correspond to Mueller matrices; we highlight the extra freedom allowed by quantum mechanics as well as the mutual constraints between the classical and quantum descriptions \add{that can now be elucidated}. \add{For example, classical polarimetry limits the quantum operations to linear transformations of Stokes operators, and classical depolarization is in turn limited to photon-number-nonconserving quantum operations. This leads us to conjecture a quantum-mechanical origin for an oft-assumed tenant of classical polarimetry \cite{Cloude1986,Simon1987,Kimetal1987,Simonetal2010,GamelJames2011}. } 

\del{This }\add{Our formalism } allows us to specify the action of polarimetric elements on quantum states, \add{which continue to be optimized \cite{SarsenValagiannopoulos2019}.} \del{which }\add{This } is crucial to quantum-enhanced parameter estimation\add{, whereby the optimal measurement precision for a set of parameters can be found for a given input state and quantum operation. Specifically, quantum enhancements in the simultaneous estimation of multiple parameters exist \cite{Humphreysetal2013}, and can be used to detect classically undetectable parameters \cite{Tsangetal2016,Pauretal16,Yangetal16,Tangetal2016,Thametal2017,Rehaceketal2017,Chrostowskietal2017,Parniaketal2018}. The new framework developed here will allow for similar advantages to be obtained for shot-noise-limited polarimetry}.

\section{Background}
\subsection{Stokes parameters and degree of polarization}
Polarization characterizes the intensity of electromagnetic waves along various component axes.
The simplest example is a monochromatic electromagnetic wave propagating in direction $\mathbf{k}$, which can be written as
\eq{
	\mathbf{E}\left(\mathbf{r},t\right)=\left(\boldsymbol{h} h+\boldsymbol{v}v\right)\eu^{\iu \mathbf{k}\cdot\mathbf{r}-\iu\omega t},
	\label{eq:plane wave E field}
}
for complex constants $ h$ and $ v$ and orthogonal polarization vectors $\boldsymbol{h}$ and $\boldsymbol{v}$ that are transverse to the propagation direction $\mathbf{k}$. The polarization properties of any field $\mathbf{E}$ are characterized by the four Stokes parameters, which measure the total intensity as well as the intensity differences along various axes:
\eq{
	S_{0,\text{mono}}&=\left|\boldsymbol{h}\cdot\mathbf{E}\right|^2+\left|\boldsymbol{v}\cdot\mathbf{E}\right|^2=\left| h\right|^2+\left| v\right|^2\\
	S_{1,\text{mono}}&=\left|\boldsymbol{h}\cdot\mathbf{E}\right|^2-\left|\boldsymbol{v}\cdot\mathbf{E}\right|^2=\left(\left| h\right|^2-\left| v\right|^2\right)\\
	S_{2,\text{mono}}&=2\Re\left[\left(\boldsymbol{h}\cdot\mathbf{E}\right)^*\left(\boldsymbol{v}\cdot\mathbf{E}\right)\right]=\left( h^* v+ h v^*\right)\\
	S_{3,\text{mono}}&=2\Im\left[\left(\boldsymbol{h}\cdot\mathbf{E}\right)^*\left(\boldsymbol{v}\cdot\mathbf{E}\right)\right]=-\iu\left( h^* v- h v^*\right).
	\label{eq:classical Stokes parameters}
}
There are only three free parameters in the Stokes parameters for a plane wave given by \eqref{eq:plane wave E field}, as the Stokes parameters satisfy the identity $S_{0,\text{mono}}^2=S_{1,\text{mono}}^2+S_{2,\text{mono}}^2+S_{3,\text{mono}}^2$. This leads to the definition of a vector
$\mathbf{S}_{\text{mono}}\equiv\left(S_{1,\text{mono}},S_{2,\text{mono}},S_{3,\text{mono}}\right)$ that, when normalized by $S_0$, spans a unit sphere known as the Poincar\'e sphere \cite{Collett1992}. The angular coordinates of this vector and the total intensity $S_0$ encompass all of the polarization information of a plane wave.

Quasi-monochromatic or stochastic light requires taking time or ensemble averages of \eqref{eq:classical Stokes parameters} $S_\mu=\left\langle S_{\mu,\text{mono}}\right\rangle_\text{classical}$, with the vector $\mathbf{S}/S_0$ in general lying inside of the Poincar\'e sphere. This allows us to define the degree of polarization \cite{Wiener1930} 
\eq{p=\frac{\left|\mathbf{S}\right|}{S_0}\label{eq:p degree of polarization}.} 
 All classical beams of light can be written as unique sums of a perfectly polarized and a completely unpolarized beam, and the degree of polarization $p$ quantifies the relative contributions of the two \cite{Wolf1959,BornWolf1999,Collett1992}.

Quantizing the electromagnetic field $\mathbf{E}$ inside a volume $V$ leads to the quantization rules
 $ h\to\sqrt{\frac{\hbar\omega}{2V\epsilon_0}}\had_H$ and $ v\to\sqrt{\frac{\hbar\omega}{2V\epsilon_0}}\had_V$ , where the operators obey bosonic commutation relations $\left[\ha_i,\had_j\right]=\delta_{ij}$
 . For notational consistency we choose to discuss annihilation operators in the circularly polarized basis $\haL\equiv \left(\ha_H-\iu\ha_V\right)/\sqrt{2}$ and $\haR\equiv \left(\ha_H+\iu\ha_V\right)/\sqrt{2}$. This leads to the Stokes operators defined by \cite{Fano1949,Fano1954,JauchRohrlich1955} \eq{
	&\hat{S}_0=\haLd\haL+\haRd\haR \\
	&\hat{S}_1=\haLd\haR+\haRd\haL 
	\\
	&\hat{S}_2=-\iu\left(\haLd\haR-\haRd\haL\right) 
	\\
		&\hat{S}_3=\haLd\haL-\haRd\haR 
	\label{eq:Quantum Stokes operators},
}
which satisfy the $\mathfrak{su}$(2) algebra (up to a factor of 2)
\eq{
	\left[\hat{S}_\mu,\hat{S}_\nu\right]&=2\iu\left(1-\delta_{\mu 0}\right)\left(1-\delta_{\nu 0}\right)\sum_{j=1}^3\epsilon_{\mu\nu j} \hat{S}_j\\
	\hat{S}_1^2&+\hat{S}_2^2+\hat{S}_3^3=\hat{S}_0^2+2\hat{S}_0,
}
and are related to the classical Stokes parameters by $S_\mu=\left\langle\hat{S}_\mu\right\rangle$, where $\left\langle\bullet\right\rangle$ denotes the quantum expectation value
\cite{JauchRohrlich1955,Collett1970}. Since the Stokes operators do not commute, there is a diverse set of quantum states with identical classical polarization properties. The decomposition of general quantum states into perfectly polarized and completely \add{un}polarized components still exists, but is no longer unique \cite{GoldbergJames2017}. \add{For example, the classical two-mode coherent states $\ket{\alpha}_L\otimes\ket{\beta}_R$ that are routinely generated using lasers are perfectly polarized, but so too are the $N$-photon projections of these states \cite{AtkinsDobson1971,Luis2016}. } This diversity has inspired new definitions for the degree of polarization \cite{AlodjantsArakelian1999,Luis2002,Klimovetal2005,SanchezSotoetal2006,Luis2007PRAtypeII,Luis2007OptComm,Klimovetal2010,Bjorketal2010}, and may lead to technological advantages \cite{Bjorketal2015,Bjorketal2015PRA,Bouchardetal2017,GoldbergJames2018Euler}.

\subsection{Mueller matrices: transformations of Stokes parameters}
Polarimetry characterizes materials by how they linearly transform the polarization properties of incident light. In general, all four Stokes parameters can change, leading to the transformation
\eq{S_\mu\to \sum_{\nu=0}^{3}M_{\mu\nu}S_\nu,\label{eq:Stokes Mueller transform}} where 
the $4\times4$ matrix $M$ is termed the Mueller matrix. The goal of polarimetry is to find the components of $M$ by shining light with known Stokes parameters on a material and measuring the Stokes parameters of the output light \add{\cite{Hauge1980}}. \add{This is essential for transmission ellipsometry, in which one seeks to compare the change in polarization $M$ to a modelled transformation for light travelling through a given medium \cite{AzzamBashara1977}.}

The uncertainty of the estimated components of $M$ is typically optimized by repeatedly shining bright, perfectly polarized light (large $S_0$, $p=1$) on an object, using a different polarization orientation for each repetition, and then using matrix inversion to determine the Mueller matrix \cite{Laydenetal2012}. \add{Classically, the four Stokes parameters can be simultaneously estimated \cite{Azzam1985}, while the analogous quantum Stokes operators do not commute and thus cannot be simultaneously estimated. } However, \del{this }\add{the }classical scheme\add{s } may be outdone by quantum-inspired ones; recent work implies that certain types of Mueller matrices are more efficiently estimated using \textit{unpolarized} light \cite{GoldbergJames2018Euler}. In this quantum scheme, multiple components of the Mueller matrix may be \textit{simultaneously} estimated with better-than-classical precision. \add{Because the components of the Mueller matrix are estimated directly, as opposed to an experimenter attempting to measure the non-commuting Stokes operators and then performing a matrix inversion, it is possible to simultanously estimate multiple parameters of a Mueller matrix in quantum-mechanical scenarios. This is in contrast to earlier quantum-mechanical strategies that avoid the non-commutability problem by using ensembles of single photons \cite{Lingetal2006}, and directly makes use of the quantum properties of states of light that are more sophisticated than identical copies of photons. } We seek to provide a connection between Mueller matrices and quantum operations such that similar quantum-inspired schemes can be provided for all of polarimetry.

Mueller matrices are broadly categorized as depolarizing or nondepolarizing. Nondepolarizing matrices are those that permit a description in terms of Jones matrices via an SL($2,\mathds{C}$) transformation on the electric field spinor from \eqref{eq:plane wave E field}: \eq{\mathbf{E}\propto\begin{pmatrix}
		h\\ v
	\end{pmatrix}\to \begin{pmatrix}
		J_{11}&J_{12}\\J_{21}&J_{22}
	\end{pmatrix}\begin{pmatrix}
		h\\ v
	\end{pmatrix}\equiv\mathbf{J}\mathbf{E},} while depolarizing matrices are necessarily ensemble averages of their nondepolarizing counterparts \cite{Kimetal1987}:
\eq{\Phi=\mathbf{E}\mathbf{E}^\dagger\to\sum_i \lambda_i \mathbf{J}_i\Phi\mathbf{J}_i^\dagger.}
The former do not change the degree of polarization of perfectly polarized incident light and the latter reduce its polarization. This nomenclature is somewhat misleading, however, as partially polarized light can have its degree of polarization both increase and decrease under the effect of both depolarizing and nondepolarizing optical systems \cite{Simon1990}. An alternative nomenclature terms nondepolarizing systems as deterministic, due to their realizability as pure Jones matrices, and depolarizing systems are deemed nondeterministic, due to their realizability as ensemble averages of Jones operations \cite{Simon1990}. This categorization has an important parallel for quantum systems.

Mueller matrices for elements commonly used to control polarization are tabulated in various sources (see, e.g., \add{\cite{AzzamBashara1977}}\cite{Chipman1995} and references therein). These include  retarders, which maintain the intensity $S_0$ and degree of polarization $p$ while rotating the polarization vector $\mathbf{S}$, making them deterministic/nondepolarizing:
\eq{M_R=\begin{pmatrix}
		1&\mathbf{0}\\
		\mathbf{0}^T&\mathbf{R}
	\end{pmatrix}} for three-dimensional rotation matrix $\mathbf{R}$;
 diattenuators, which differentially transmit light incident with different polarization directions while taking perfectly polarized light to perfectly polarized light at a reduced intensity, and are likewise deterministic/nondepolarizing:
 \eq{\text{e.g.,}\, 
 	M_D=\begin{pmatrix}
 		\tfrac{q+r}{2}&\tfrac{q-r}{2}&0&0\\
 		\tfrac{q-r}{2}&\tfrac{q+r}{2}&0&0\\
 		0&0&\sqrt{qr}&0\\
 		0&0&0&\sqrt{qr}
 	\end{pmatrix}, 
 \quad 0\leq q,r\leq 1;\label{eq:Diattenuator matrix example}}
 and depolarizers, which maintain the total intensity $S_0$ while reducing $p$ of perfectly polarized light, such as
 \eq{M_d=\begin{pmatrix}
 		1&\mathbf{0}\\
 		\mathbf{0}^T&\mathbf{m}
 	\end{pmatrix},\label{eq:depolarization Mueller symmetric}} for $3\times3$ symmetric matrices $\mathbf{m}$ with eigenvalues between $-1$ and $1$
 \cite{Chipman1995}. Setting $r=0$ and $q=1$ in $M_D$, for example, yields a Mueller matrix for the commonly-used linear polarizer.

An arbitrary Mueller matrix characterizing a given material can be decomposed into the sequential application of a diattenuator, a retarder, and a depolarizer:\eq{M=M_dM_DM_R.} This decomposition is unique for a given ordering of the three components \cite{LuChipman1996}. Similarly, an arbitrary Mueller matrix can be decomposed into a positive sum of no more than four nondepolarizing elements \cite{Cloude1986}.
Multiplication of Mueller matrices is physically interpreted as the sequential application of the associated optical elements, and convex sums of Mueller matrices as the application of spatially or temporally varying optical elements \cite{Giletal2013}.
 A quantum description of polarimetry can thus be obtained from quantum descriptions of these optical elements, and an arbitrary Mueller matrix can be obtained by applying the associated quantum operations in the correct sequence or combination.

We briefly mention the physical constraints on the 16 independent components of Mueller matrices (this subject has a significant history \cite{FryKattawar1981,GilBernabeu1985,Cloude1986,Barakat1987,Cloude1990,GivensKostinski1993,Nagirner1993,vanderMee1993,Raoetal1998,Gil2000,vanZyletal2011,Giletal2013,Zakerietal2013}). The chief physical requirement is that Mueller matrices take all physically-viable sets of Stokes parameters to physically-viable sets of Stokes parameters ($p\leq 1$). Further, Mueller matrices must always be Hermitian under a specific change of basis \cite{Simon1982}, and the associated Hermitian matrices should be positive \cite{GilBernabeu1985,Cloude1986,Gil2000} (quantum-inspired justifications are given in Refs. \cite{Simonetal2010,GamelJames2011}). We will see that there exist diverse sets of quantum operations that both do and do not satisfy these constraints.

\subsection{SU(2), Quantum Stokes Operators, and Representations of Rotations}
\add{The $\mathfrak{su}\left(2\right)$ algebra mathematically describes $N$-qubit states, which include photons, electrons, and other spin-$N/2$ systems, allowing results obtained in one physical system to be applied to the rest. }
\del{There are many }\add{Many }properties \del{about }\add{of } the $\mathfrak{su}\left(2\right)$ algebra \del{that }play a role in the quantum description of polarimetry. The most important is that $\hat{S}_0$ commutes with the other Stokes operators, and thus the relevant Hilbert space is a tensor sum of subspaces with different photon numbers $N$. This ensures that the Stokes parameters never contain any information about superpositions of states with different photon numbers. \add{To wit, quantum coherence between states of light with different numbers of photons is guaranteed to have zero effect on polarization.}

SU(2) operations can represented by a variety of triplets of parameters. For example, 
\eq{\hat{R}=\eu^{-\iu\tfrac{\phi}{2}\hat{S}_z}\eu^{-\iu\tfrac{\theta}{2}\hat{S}_y}\eu^{-\iu\tfrac{\psi}{2}\hat{S}_z}=\exp\left(-\iu\frac{\chi }{2}\mathbf
	{\hat{S}}\cdot\mathbf{n}\right)}
defines a rotation by a set of Euler angles $\phi\in \left(0,2\pi\right)$, $\theta\in \left(0,\pi\right)$, and $\psi\in \left(0,2\pi\right)$, or by a rotation angle $\chi\in\left(0,2\pi\right)$  for a counter-clockwise rotation around the axis in the $\Theta\in\left(0,\pi\right)$, $\Phi\in\left(0,2\pi\right)$-direction specified by unit vector $\mathbf{n}=\left(\sin\Theta\cos\Phi,\sin\Theta\sin\Phi,\cos\Phi\right)$. Relationships between these sets of angles as well as numerous other properties are tabulated \cite{Varshalovichetal1988}; note that one must include the normalization factor of $\tfrac{1}{2}$ for \eqref{eq:Quantum Stokes operators} to agree with the usual SU(2) notation. These rotations have the effect of transforming the Stokes operators as vectors under three-dimensional rotations:
\eq{\hat{R}\mathbf{\hat{S}}\hat{R}^\dagger=\mathbf{R}^T\mathbf{\hat{S}},} where $\mathbf{R}$ is a $3\times 3$ rotation matrix that rotates a vector counter-clockwise in accordance with the relevant Euler-angle or angle-axis prescription. The rotation operators also generate linear transformations of the creation and annihilation operators.

Some simple examples suffice to generate the pertinent rotation operation results.
A rotation around the $z$-axis yields \cite{Leonhardt2003} \eq{\eu^{-\iu\psi S_z/2}\begin{pmatrix}
S_x\\S_y\\S_z
\end{pmatrix}\eu^{\iu\psi S_z/2}=\begin{pmatrix}
\cos\psi&\sin\psi&0\\-\sin\psi&\cos\psi&0\\0&0&1
\end{pmatrix}\begin{pmatrix}
S_x\\S_y\\S_z
\end{pmatrix},
\label{eq:z axis rotation}} and the rest of the rotation matrices follow from cyclic permutations of the operators as well as composing rotation operations. \add{The rotation given by \eqref{eq:z axis rotation} is equivalent to applying a relative phase between two orthogonal modes $L$ and $R$, similar to the relative phases that can be applied between a pair of spatial modes \cite{Spechtetal2009}, and can be done in a controlled manner \cite{Becketal2016}. } According to the Euler-angle parametrization, we \del{thus } have
\begin{widetext}\eq{\mathbf{R}^T=\left(
		\begin{array}{ccc}
			\cos (\theta ) \cos (\phi ) \cos (\psi )-\sin (\phi ) \sin (\psi ) & \cos (\psi ) \sin (\phi )+\cos (\theta ) \cos (\phi ) \sin (\psi ) & -\cos (\phi ) \sin (\theta ) \\
			-\cos (\theta ) \cos (\psi ) \sin (\phi )-\cos (\phi ) \sin (\psi ) & \cos (\phi ) \cos (\psi )-\cos (\theta ) \sin (\phi ) \sin (\psi ) & \sin (\theta ) \sin (\phi ) \\
			\cos (\psi ) \sin (\theta ) & \sin (\theta ) \sin (\psi ) & \cos (\theta ) \\
		\end{array}
		\right)\label{eq:rotation matrix euler}}\end{widetext} and \eq{\begin{pmatrix}
\haLd\\\haRd
\end{pmatrix}\to \left(
\begin{array}{cc}
e^{-\frac{1}{2} i (\phi +\psi )} \cos \left(\frac{\theta }{2}\right) & e^{-\frac{1}{2} i (\phi -\psi )} \sin \left(\frac{\theta }{2}\right) \\
-e^{\frac{1}{2} i (\phi -\psi )} \sin \left(\frac{\theta }{2}\right) & e^{\frac{1}{2} i (\phi +\psi )} \cos \left(\frac{\theta }{2}\right) \\
\end{array}
\right)\begin{pmatrix}
\haLd\\\haRd
\end{pmatrix}. } These will be essential in connecting quantum operations to Mueller matrix transformations.

\section{Quantum operations for Mueller matrices}
It is now time to derive quantum operations for arbitrary Mueller matrices. Unitary quantum operations take states $\hat{\rho}\to \mathcal{E}\left(\hat{\rho}\right)=\hat{U}\hat{\rho} \hat{U}^\dagger$, enabling the transformations $\hat{S}_\mu\to \hat{U}^\dagger \hat{S}_\mu \hat{U}$. General quantum channels are represented by completely-positive and trace-preserving (CPTP) maps $\mathcal{E}\left(\hat{\rho}\right)=\sum_l \hat{K}_l\hat{\rho} \hat{K}_l^\dagger$ that transform the Stokes operators according to
\eq{\hat{S}_\mu\to \sum_l \hat{K}_l^\dagger \hat{S}_\mu \hat{K}_l,\quad \sum_l \hat{K}_l^\dag \hat{K}_l=\id.\label{eq:Kraus map}} The wide variety of suitable Kraus operators $\hat{K}_l$ allows for more diverse transformations than those represented by Mueller matrices. Importantly, the requirement that \eqref{eq:Stokes Mueller transform} be valid for the expectation values $\text{Tr}\left(\hat{\rho}\hat{S}_\mu\right)$  regardless of quantum state $\hat{\rho}$ implies that the Stokes operators themselves must transform in the same way as the Stokes parameters, via \eq{\sum_l \hat{K}_l^\dagger \hat{S}_\mu \hat{K}_l=\sum_{\nu}M_{\mu\nu}\hat{S}_\nu,} which dramatically limits the number of quantum operations that can be represented classically. We here focus on those specific operations, leaving the added richness of quantum operations to future study.

The Kraus operator map represented by \eqref{eq:Kraus map} can be physically interpreted as the action of adding an auxiliary system $v$ in its vacuum state to $\hat{\rho}$, performing a unitary operation on the enlarged system $\hat{\rho}\ot \hat{0}_v$, and then ignoring the state of the auxiliary system. We will show in the following sections that photon-number-conserving unitary operations on $\hat{\rho}$ correspond to retarders, photon-number-conserving unitary operations on $\hat{\rho}\ot\hat{0}_v$ correspond to any combination of retarders and diattenuators, and that depolarization can only be described by operations on $\hat{\rho}\ot\hat{0}_v$ that do not conserve photon number. In other words, photon-number-conserving unitary operations on $\hat{\rho}\ot\hat{0}_v$ can describe all nondepolarizing transformations, and thus describe all of Jones calculus.

Just like in classical polarimetry \add{\cite{Giletal2013}}, quantum channels can be composed of products or sums of quantum channels. The application of Mueller matrix $M_1$ followed by Mueller matrix $M_2$ corresponds to the composite channel \eq{\mathcal{E}_{M_2M_1}\left(\hat{\rho}\right)=\mathcal{E}_{M_2}\left[\mathcal{E}_{M_1}\left(\hat{\rho}\right)\right],} and a linear combination of $M_1$ and $M_2$ corresponds to the channel \eq{\mathcal{E}_{p_1M_1+p_2M_2}\left(\hat{\rho}\right)=p_1\mathcal{E}_{M_1}\left(\hat{\rho}\right)+p_2\mathcal{E}_{M_2}\left(\hat{\rho}\right).} This means that characterizing the channels corresponding to each type of Mueller matrix is sufficient for characterizing all general Mueller matrices.

\subsection{Retarders and rotation operators}
Representing Mueller matrices for retarders $M_R$ by quantum operations is achieved using the rotation operators $\hat{R}$ from above. These operators are manifestly unitary, and their transposes are also rotation operators, obeying $\hat{R}^T\left(\phi,\theta,\psi\right)=\hat{R}\left(\psi,-\theta,\phi\right)$. A rotation operation applied to an arbitrary quantum state $\hat{\rho}\to \hat{R}^T\hat{\rho}\hat{R}^*$ immediately generates the transformation
\eq{
\begin{pmatrix}
	\hat{S}_0\\
	\hat{S}_1\\
	\hat{S}_2\\
	\hat{S}_3
\end{pmatrix}\to\hat{R}^*\begin{pmatrix}
\hat{S}_0\\
\hat{S}_1\\
\hat{S}_2\\
\hat{S}_3
\end{pmatrix}\hat{R}^T=\begin{pmatrix}
1&\mathbf{0}\\
\mathbf{0}^T&\mathbf{R}
\end{pmatrix}\begin{pmatrix}
\hat{S}_0\\
\hat{S}_1\\
\hat{S}_2\\
\hat{S}_3
\end{pmatrix}=M_R\begin{pmatrix}
\hat{S}_0\\
\hat{S}_1\\
\hat{S}_2\\
\hat{S}_3
\end{pmatrix}.
} The three Mueller matrix parameters encoded in $M_R$ are precisely the three SU(2) parameters in $\hat{R}$. \add{This is our first example showing that the transformation of SU$(2)$ operators is directly responsible for classical polarization transformations.}

This transformation preserves the commutation relations of $\haL$ and $\haR$ since it transforms them by a unitary matrix $\mathbf{U}$:
\eq{\begin{pmatrix}
		\haL\\
		\haR
\end{pmatrix}\to\hat{R}^\dag \begin{pmatrix}
\haL\\
\haR
\end{pmatrix}\hat{R}=\mathbf{U}\begin{pmatrix}
\haL\\
\haR
\end{pmatrix}.} No photons are lost or gained in this transformation, as each transformed creation operator remains normalized. 

\subsection{Diattenuators and rotation operations in larger Hilbert spaces}
Diattenuation occurs when different components of the electric field are differentially transmitted by an optical device. For a classical state such as \eqref{eq:plane wave E field}, the transformation $ h\to \sqrt{q} h$ and $ v\to \sqrt{r} v$ represents transmission probabilities $q$ and $r$ for horizontally- and vertically-polarized light, respectively. This classical transformation is represented by the diattenuator Mueller matrix $M_d$ exemplified in \eqref{eq:Diattenuator matrix example}.

Diattenuation along a different axis can be parametrized by a rotation followed first by a linear diattenuation and then by the inverse of the original rotation. These rotations need only be parametrized by two angles because only the direction of the diattenuation axis can be varied, and there is no need to rotate the coordinate system about that axis. The four parameters of a general Mueller matrix representing diattenuation are thus the diattenuation strengths $q$ and $r$ as well as the two angular coordinates of the diattenuation axis. 

The quantum analog $\haL\to \sqrt{q}\haL$ and $\haR \to \sqrt{r}\haR$ immediately yields a Mueller matrix of the same form as in the classical case, this time with diattenuation of left- and right-circularly polarized light due to the definitions of $\haL$ and $\haR$, but does not preserve the commutation relations of $\haL$ and $\haR$. This prompts using an enlarged Hilbert space to represent this quantum-mechanical transformation.

\begin{figure}
	\includegraphics[width=\columnwidth,trim=0 0 3cm 0,clip]{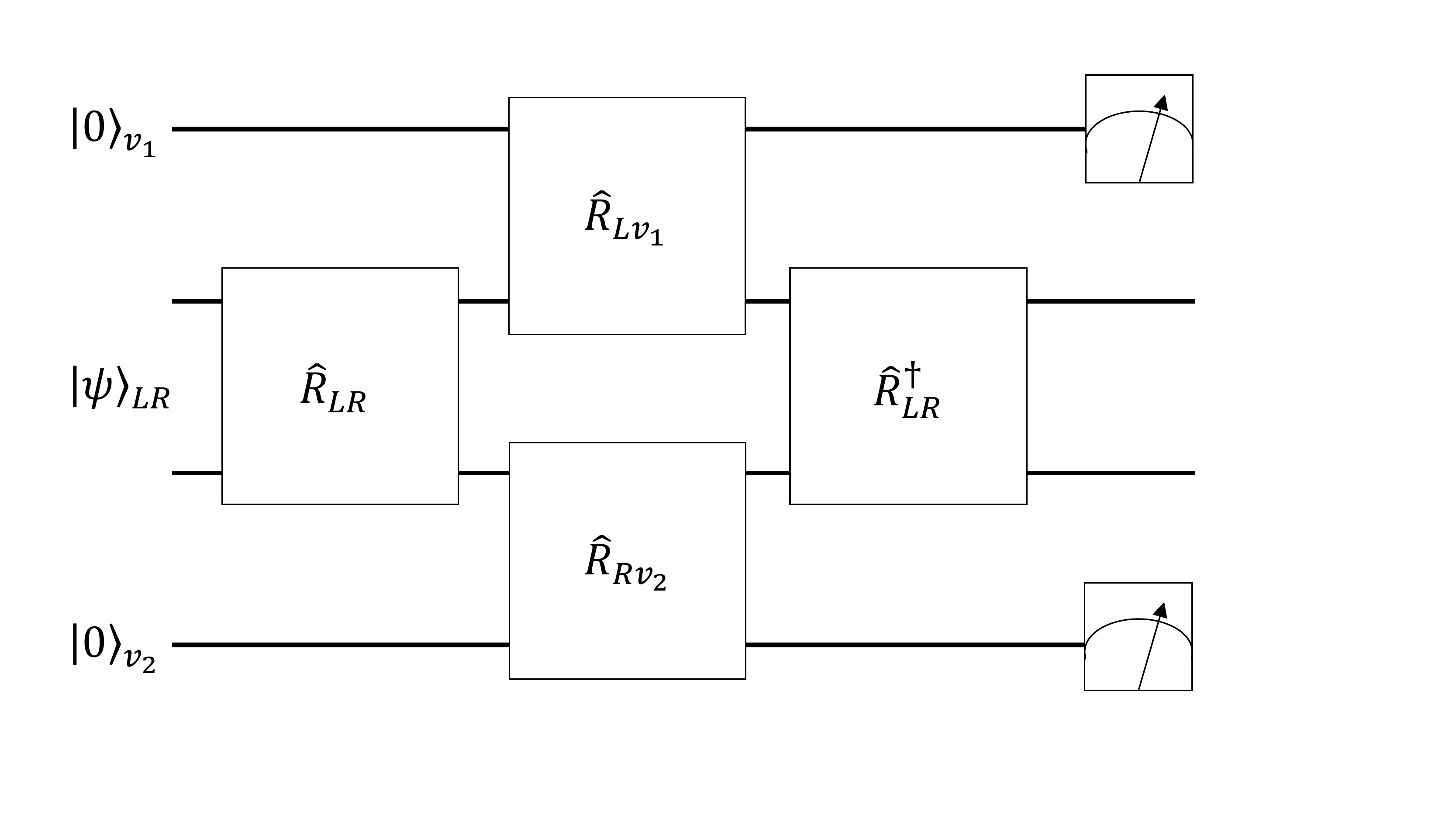}
	\caption{Quantum-circuit schematic of depolarization. A quantum state $\ket{\psi}_{LR}$ first has its polarization rotated, the two modes each lose population by being coupled to a vacuum mode that is then neglected, and the state is rotated back to its original basis, with the arguments of the rotation operators in the diagram  given in  \eqref{eq:diattenuation unitary 1} in terms of Euler angles. The entire circuit corresponds to the transformation \eqref{eq:diattenuation on ab} and transforms the Stokes vectors via a diattenuation Mueller matrix for arbitrary input states $\ket{\psi}_{LR}$.}
	\label{fig:depolarization two vac}
\end{figure}

One method is to add an auxiliary vacuum mode to both $L$ and $R$, perform a rotation between each mode and its corresponding vacuum mode, and then trace over the auxiliary modes. The joint operator \eq{\hat{U}_D\equiv&\hat{R}_{L,R}^\dag\left(0,\theta,\psi\right) \hat{R}_{L,\text{vac}_1}\left(0,2\cos^{-1}\sqrt{q},0\right)\\&\quad\times\hat{R}_{R,\text{vac}_2}\left(0,2\cos^{-1}\sqrt{r},0\right)\hat{R}_{L,R}\left(0,\theta,\psi\right)\label{eq:diattenuation unitary 1}} acting on the enlarged state $0_{\text{vac}_1}\otimes\hat{\rho}\otimes0_{\text{vac}_2}$ yields the desired transformation of $\haL$ and $\haR$ in a trace-preserving fashion after tracing out the two vacuum modes (all arguments given in terms of Euler angles). 
We depict this method schematically in Fig. \ref{fig:depolarization two vac}, achieving the transformation
\begin{widetext}\eq{
	\begin{pmatrix}
	\haLd\\
	\haRd
\end{pmatrix}\to
		\frac{1}{2}
		\begin{pmatrix}
			 \sqrt{q}+\sqrt{r}+\left(\sqrt{q}-\sqrt{r}\right) \cos \left(\theta \right) &  \eu^{\iu \psi } \left(\sqrt{q}-\sqrt{r}\right) \sin \left(\theta \right)\\
			 \eu^{-\iu \psi } \left(\sqrt{q}-\sqrt{r}\right) \sin \left(\theta \right) & \sqrt{q}+\sqrt{r}+\left(\sqrt{r}-\sqrt{q}\right) \cos \left(\theta \right)
		\end{pmatrix}
	\begin{pmatrix}
		\haLd\\
		\haRd
	\end{pmatrix}
\label{eq:diattenuation on ab}
	,}\end{widetext} from which the elements of Mueller matrix can easily be obtained (see Appendix \ref{app:explicit results} for explicit results).

\begin{figure}
	\includegraphics[width=\columnwidth,trim=4cm 3cm 4cm 3cm,clip]{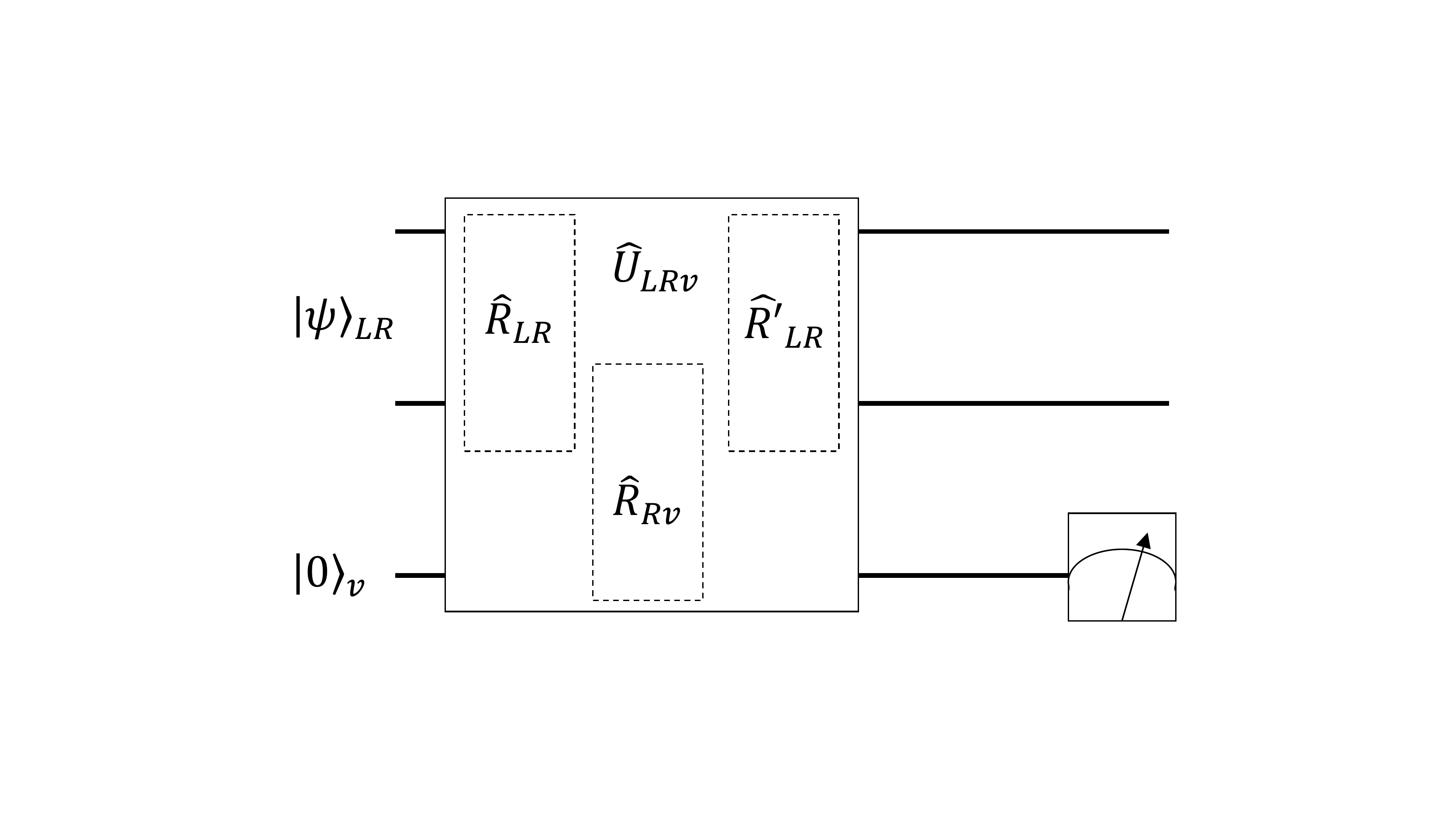}
	\caption{Quantum-circuit diagram of arbitrary combinations of depolarization and rotations. A quantum state $\ket{\psi}_{LR}$ and an auxiliary vacuum state evolve according to an SU(3) transformation $\hat{U}_{LRv}$, which contains all 7 degrees of freedom of a nondepolarizing Mueller matrix plus an additional phase.
The entire circuit corresponds to the transformation \eqref{eq:diattenuator and retarder SU3} and transforms the Stokes vectors via a nondepolarizing Mueller matrix for arbitrary input states $\ket{\psi}_{LR}$. It can alternatively be decomposed into the product of three rotation operations as per \cite{Recketal1994}, with the first and last Euler angle of the middle rotation being the same \cite{deGuiseetal2018}.
}
	\label{fig:depolarization one vac}
\end{figure}

Another method is to use a single auxiliary vacuum mode $\hat{v}$, and to perform an SU(3) operation $\hat{U}_{LRv}$ on the combined system (see Fig. \ref{fig:depolarization one vac}):
\eq{\hat{\rho} \ot&\hat{0}_v\to\hat{U}_{LRv}\hat{\rho} \ot\hat{0}_v\hat{U}_{LRv}^\dagger
	\\&
	\Rightarrow\begin{pmatrix}
		\haL\\\haR\\\hat{v}
	\end{pmatrix}\to 
	\hat{U}_{LRv}\begin{pmatrix}
		\haL\\\haR\\\hat{v}
	\end{pmatrix}\hat{U}_{LRv}^\dag=
	\mathbf{U}_{LRv}\begin{pmatrix}
		\haL\\\haR\\\hat{v}
	\end{pmatrix}.\label{eq:diattenuator and retarder SU3}} The $3\times 3$ unitary matrix $\mathbf{U}_{LRv}$ automatically preserves photon number. If we choose its upper-left $2\times 2$ block to be the matrix elements in \eqref{eq:diattenuation on ab} then $\hat{U}_{LRv}$ corresponds to pure diattenuation, with four degrees of freedom available in the remaining elements of $\mathbf{U}_{LRv}$. This is our first example of quantum degrees of freedom that have no effect on the classical polarization behaviour. 

If we allow for arbitrary $\hat{U}_{LRv}$, 
then this transformation encompasses all nondepolarizing Mueller matrices. The upper-left $2\times 2$ block of $\mathbf{U}_{LRv}$ has four complex degrees of freedom, corresponding to the four complex elements of a pure Jones transformation matrix, equivalent to the eight real degrees of freedom of SU(3) operators. The corresponding Mueller matrices have only seven real parameters because the Mueller matrices cannot encapsulate absolute phase, neglecting the global phase of the upper-left $2\times 2$ block of $\mathbf{U}_{LRv}$. Many subsequent reparametrizations of $\hat{U}_{LRv}$ in terms of a series of rotation operations exist \cite{Recketal1994,Clementsetal2016,deGuiseetal2018}; one is depicted in Fig. \ref{fig:depolarization one vac}. These together encompass all nondepolarizing Mueller matrices.

The unitary operations described by \eqref{eq:diattenuator and retarder SU3} account for all deterministic Mueller matrices: arbitrary combinations of the three parameters of retarders and the four parameters of diattenuators. They can be recast into a Kraus operator representation of a quantum channel acting only on $\hat{\rho}$ of the form \eqref{eq:Kraus map} with some minimum number of Kraus operators, such as through $\hat{K}_l=\bra{l}_v\hat{U}_{LRv}\ket{0}_v$. While varying the number of Kraus operators represents some freedom in a description of the quantum channel, even the decomposition with the minimum number of Kraus operators is only unique up to unitary transformations $\hat{K}_{l^\prime}\to\sum_{n}u_{l^\prime l}\hat{K}_l$ for some unitary operator with matrix elements $u_{l^\prime l}$. We stress that this freedom underlies only the \textit{description} of the quantum channel for the classical Mueller matrix; the \textit{effect} of the quantum channel is always the same as that of \eqref{eq:diattenuator and retarder SU3}. Because of this we conclude that all \add{classical }Mueller matrices that can be described by pure nondepolarizing operations correspond to photon-number-conserving \add{quantum }operations within a potentially-enlarged Hilbert space.

It is intriguing that the classical nondepolarizing degrees of freedom contained in $\mathbf{J}\in\text{SL}\left(2,\mathds{C}\right)$ are here recast into degrees of freedom of $\mathbf{U}_{LRv}\in\text{SU}(3)$. 
These groups certainly do not have the same structure, yet conspire to yield the same physical transformations.
Our results show that any Jones matrix $\mathbf{J}$ acting on $\mathbf{E}$ can be represented as a truncated unitary matrix acting on a vector of creation operators $\left(\haLd,\haRd,\cdots\right)^T$ if the remaining modes begin in their vacuum states. This truncation always yields a complex $2\times 2$ matrix that can be rescaled to become an element of $\text{SL}(2,\mathds{C})$. Restricting to a single extra mode, such a truncation is equivalent to projecting from points on a hypersphere $S^5$ to points on $S^3$ \cite{ZyczkowskiSommers2000}. These truncations have been studied from the point of view of random matrices, for which the distribution of points as well as the distribution of the eigenvalues of the truncated unitaries are known analytically \cite{ZyczkowskiSommers2000}, and may be important for predicting the effect of a random nondepolarizing Mueller matrix.

\subsection{Depolarizers and number-non-conserving operations}
Seeing that number-conserving quantum operations parametrize the seven parameters of nondepolarizing Mueller matrices, the remaining nine free parameters of a general Mueller matrix must be described by quantum operations that do not conserve photon number in an enlarged Hilbert space. One must instead allow for unitary operations that change the total number of excitations in the larger space. As shown explicitly in \cite{MoyanoFernandezGarciaEscartin2017}, unitary transformations on a state allow one to generate a much larger set of transformations than unitary transformations on creation operators . 

Quantum theories of depolarization have been studied \cite{Ramos2005,doNascimentoRamos2005,BallBanaszek2005,Klimovetal2008,RefregierLuis2008,KlimovSanchezSoto2010,RivasLuis2013}. Depending on the desired properties, one can compose various dynamical equations that lead to depolarized quantum states. However, this depolarization is usually restricted to one or two of the degrees of freedom of Mueller matrices; we seek a quantum description that simultaneously encompasses all of the remaining degrees of freedom.

A typical example of a depolarization process is partial depolarization, whereby perfectly polarized light is transformed into light with degree of polarization $p$, with ideal depolarization corresponding to $p=0$:
\eq{\begin{pmatrix}
		\hat{S}_0\\
		\hat{S}_1\\
		\hat{S}_2\\
		\hat{S}_3
\end{pmatrix}\to \begin{pmatrix}
1&0&0&0\\
0&p&0&0\\
0&0&p&0\\
0&0&0&p\\
\end{pmatrix}\begin{pmatrix}
\hat{S}_0\\
\hat{S}_1\\
\hat{S}_2\\
\hat{S}_3
\end{pmatrix}.} This is normally described by a quantum channel $\hat{\rho}\to p\hat{\rho}+\left(1-p\right)\id/\mathcal{N}$ with normalization constant $\mathcal{N}=\text{Tr}(\id)$ \cite{NielsenChuang2000}. However, that transformation does not always yield $\hat{S}_0\to\hat{S}_0$, because the  probability of $\hat{\rho}$ being in a particular photon-number subspace is not guaranteed to be the same as the probability of $\id$ being in that subspace. Knowing the weight $p_N$ of $\hat{\rho}$ in each subspace would allow for an appropriate transformation \eq{\hat{\rho}\to p\hat{\rho}+\left(1-p\right)\sum_N p_N\frac{\id_N}{N+1},\label{eq:depol quantum map}} but we seek transformations that are independent of the initial state.

One way of achieving \eqref{eq:depol quantum map} is through a continuum of Kraus operators:
\eq{\hat{\rho} \to \left(\sqrt{p}\id\right) \hat{\rho}\left(\sqrt{p}\id\right)^\dag+\int \left(\sqrt{1-p}\hat{R}\right)\hat{\rho}\left(\sqrt{1-p}\hat{R}\right)^\dag d\hat{R},\label{eq:depol quantum map better}} where the integrand contains a normalized Haar measure $d\hat{R}$ for the rotation operators. This physically corresponds to the effect of random SU(2) rotations \cite{RivasLuis2013}. The validity of this transformation can be immediately verified using the properties $\int \hat{R}\hat{S}_0\hat{R}^\dag d\hat{R}=S_0$ and $\int \hat{R}\hat{S}_i\hat{R}^\dag d\hat{R}=0$ for $i=1,2,3$. If we seek a unitary operation in an enlarged Hilbert space to describe \eqref{eq:depol quantum map better}, we require a continuum of orthonormal states $\ket{\boldsymbol{\theta}}_v$ to be populated without transferring any population from the initial system:
\eq{
\hat{\rho}\ot\hat{0}_v\to \hat{U}\hat{\rho}\ot\hat{0}_v\hat{U}^\dag,\quad\hat{U}= \int \hat{R}_{LR}\left(\boldsymbol{\theta}\right)\ot \ket{\boldsymbol{\theta}}\bra{0}_v d\boldsymbol{\theta}.
} This is our first glimpse into the general feature that depolarization channels correspond to transformations that cannot be thought of as photon-number-conserving.

We next seek quantum descriptions of polarization maps described by \eqref{eq:depolarization Mueller symmetric}. Since this type of depolarization does not act isotropically on the Stokes operators, perhaps an arbitrary depolarization matrix can be constructed by adding an appropriate weight function $f\left(\hat{R}\right)$ to the integrand in \eqref{eq:depol quantum map better}. 
It turns out that Mueller matrices $M_d$ with arbitrary $\mathbf{m}$ can indeed be constructed by a weighted integral of SU(2) operations if we allow for arbitrary $f$. However, the corresponding Kraus operators $\hat{K}_l\propto\hat{R}_l\equiv\hat{R}\left(\phi_l,\theta_l,\psi_l\right)$ need to incorporate the weight factor $\sqrt{f}$; this restricts $f$ to the set of positive-semidefinite functions, which are not sufficient for constructing all $\mathbf{m}$ while ensuring $\hat{S}_0\to\hat{S}_0$.

For example, take the rotation operation decomposed into the three Euler angles as per \eqref{eq:rotation matrix euler} and define the weight function $f\left(\phi,\theta,\psi; a,b,c,d,e,f,g,h,i,j\right)$:
	\eq{
		f=&\left[-4 a \sin (\psi) \sin (\phi)+4 b \sin (\psi) \cos (\phi)-\pi  c \cos (\psi)
		\right.\\ &\left.
		-4 d \cos (\psi) \sin (\phi)+4 e \cos (\psi) \cos (\phi)+\pi  f \sin (\psi)
		\right.\\ &\left.
		+\pi  g \cos (\phi)+\pi  h \sin (\phi)+2 i \cos (\theta)+j\right]/4\pi^3
		.}
If we evaluate the effect of the transformation $\int_0^{2\pi}d\phi\int_0^{\pi}d\theta\int_0^{2\pi}d\psi\, f\hat{R}\hat{\rho}\hat{R}^\dag$ (in which the measure differs from the Haar measure by a factor of $\sin\theta$),
we find the Mueller matrix
\eq{M=\begin{pmatrix}
		j&0&0&0\\
		0&a&b&c\\
		0&d&e&f\\
		0&g&h&i\\
\end{pmatrix},} which for $j=1$, $b=d$, $c=g$, $f=h$ describes a general depolarizer. However, $f$ is clearly negative for some values of its argument, meaning this does not describe a valid Kraus operator map.

In general there are quite a few weight functions that allow for arbitrary parameters $a$ through $i$, but finding ones that are normalized ($j=1$) without compromising the independence of the other parameters is challenging. Here, the requirement of a CPTP map enforces the same constraint as derived classically for Mueller matrices. We see that CPTP maps generating Mueller matrices with $M_{00}=1$ can be formed by positive combinations of SU(2) rotations.
Correspondingly, the only classically valid Mueller matrices with $M_{00}=1$ are positive sums of nondepolarizing Mueller matrices that each have $M_{00}=1$ (e.g., \cite{GamelJames2011}); namely, linear combinations of rotation operations. This intimate connection strongly suggests  a quantum-mechanical origin for the requirement that depolarizing Mueller matrices be formed from positive combinations of nondepolarizing Mueller matrices \cite{Cloude1986,Gil2000,Simonetal2010,GamelJames2011}.

We have not shown that quantum operations corresponding to depolarizing Mueller matrices \textit{must} be formed from convex combinations of quantum operations corresponding to nondepolarizing Mueller matrices. While a CPTP map that enacts a combination of SU(2) operations is a \textit{sufficient} condition for generating Mueller matrices with $M_{00}=1$, it is by no means a \textit{necessary} condition. For example, given that a set of Kraus operators $\left\{\hat{K}_l\right\}$ yields a valid depolarizing Mueller matrix, so too must the set of Kraus operators $\left\{\hat{K}_{l^\prime}\right\}$, where we can choose
\eq{\hat{K}_l=\sqrt{p_l}\hat{R}_l
	,\quad \sum_l p_l=1\\
 	\hat{K}_{l^\prime}=\sum_l u_{l^\prime l}\hat{K}_l,\quad \sum_{l^\prime} u_{l^\prime k}^*u_{l^\prime l}=\delta_{kl}.
 } Each individual transformation $\hat{K}_{l^\prime}\hat{\rho}\hat{K}_{l^\prime}^\dagger$ does \textit{not} yield a Mueller matrix; in fact, the individual transformations of the Stokes operators no longer yield a linear combinations of Stokes operators. By no means does quantum mechanics mandate that a depolarizing Mueller matrix be \textit{described} by a convex combination of processes corresponding to nondepolarizing Mueller matrices.

Still, we hypothesize that CPTP maps that send Stokes operators to linear combinations of Stokes operators always \textit{permit} a decomposition into convex combinations of CPTP maps corresponding to nondepolarizing Mueller matrices. If we restrict our attention to maps that do not intermix the different photon-number subspaces, the results of \cite{GamelJames2011} immediately prove our hypothesis: since a map corresponding to a Mueller matrix must have the same behaviour on the Stokes operators regardless of the quantum state, one could work exclusively in the single-photon subspace, on which CPTP maps correspond exactly to the classical result of requiring convex combinations of nondepolarizing Mueller matrices. It is not far-fetched to believe that only those maps that do not intermix the different photon-number subspaces can send Stokes operators to linear combinations of Stokes operators. Validating this claim for general quantum operations will be the result of future work.

Convex combinations of the operations described in \eqref{eq:diattenuator and retarder SU3} are thus sufficient for describing all classical Mueller matrices. 
If the Kraus operators $\left\{\hat{K}_l^{(i)}\right\}_l$ correspond to Mueller matrix $M^{(i)}$, we have that the transformation
\eq{\mathcal{E}\left(\hat{\rho}\right)=\sum_i p_i \sum_l \hat{K}_l^{(i)}\hat{\rho} \hat{K}_l^{(i)\dagger}} corresponds to the Mueller matrix \eq{M=\sum_i p_i M^{(i)}.} The new set of Kraus operators $\left\{\sqrt{p_i}\hat{K}_l^{(i)}\right\}_{l,i}$ is sufficient for describing $M$ quantum-mechanically. The freedom within the weights $p_i$, which need number no more than four \cite{Cloude1986}, provides the remaining degrees of freedom found in arbitrary Mueller matrices. This concludes our quest for determining quantum operations corresponding to all Mueller matrices.

\subsubsection{Other descriptions of depolarizers}
Quantum mechanics may provide more insight into other forms of depolarizing matrices. 
Since the lower-right $3\times 3$ submatrix of Mueller matrices representing depolarizers must be symmetric, the remaining three parameters describing arbitrary depolarization come as the vector $\mathbf{P}$ in \eq{M_d=\begin{pmatrix}
		1&\mathbf{0}\\
		\mathbf{P}^T&\mathbf{m}
\end{pmatrix}.\label{eq:depolarizer with P}} There is no way that this can be construed as a positive sum of nondepolarizing Mueller matrices \cite{Gil2000}, but this form of depolarizers has indeed been used (e.g. \cite{LuChipman1996}).
It is hard to conceive of quantum operations that take $\hat{S}_i\to\sum_{j=1}^3M_{ij}\hat{S}_j+M_{i0}\hat{S}_0$ for nonzero $M_{i0}$ that retain $\hat{S}_0\to\hat{S}_0$, and we will show an explicit example of where this difficulty lies. 

One example of a process described by \eqref{eq:depolarizer with P} is the operation that takes \textit{all} incident light and converts it to light of a specific polarization [$\mathbf{m}=\textbf{0}$ and $\mathbf{P}=\left(0,0,1\right)$]. For left-circular polarization, this is the operation
\eq{
\mathcal{E}&\left(\hat{\rho}\right)= \\&\sum_{M=0}^\infty\sum_{m=0}^M \ket{M,0}\bra{m,M-m} \hat{\rho} \sum_{N=0}^\infty\sum_{n=0}^N \ket{n,N-n}\bra{N,0},
} where \eq{
\ket{m,n}\equiv \frac{\haLd\vphantom{a}^m\haRd\vphantom{a}^n}{\sqrt{m!n!}}\vac.} However, such a transformation is not trace-preserving, because the Kraus operators sum to $\sum_{N=0,m,n}\ket{n,N-n}\bra{m,N-m}\neq \id$. 

We can attempt to achieve a trace-preserving transformation by enacting
\begin{widetext}
\eq{
	\mathcal{E}\left(\hat{\rho}\right)= \sum_{l=1}^L\left[\sum_{M=0}^\infty\ket{M,0}\left(\sum_{m=0}^M \bra{m,M-m}\frac{\eu^{\iu \tfrac{2\pi l m}{L}}}{\sqrt{L}}\right)\right] \hat{\rho} \left[\sum_{N=0}^\infty\left(\sum_{n=0}^N \frac{\eu^{-\iu \tfrac{2\pi l n}{L}}}{\sqrt{L}}\ket{n,N-n}\right)\bra{N,0}\right].
}
\end{widetext} For a fixed maximum particle number $L-1$ this transformation indeed converts all incoming light into left-circularly-polarized light in a trace-preserving manner, due to the orthonormality condition \eq{\frac{1}{L}\sum_{l=1}^L\eu^{\iu\tfrac{2\pi l\left(m-n\right)}{L}}=\delta_{mn},\quad \left|m-n\right|<L.} However, in order to generate a Mueller matrix transformation on the Stokes \textit{operators}, this transformation must hold for arbitrarily large photon numbers. If we allow the formal limit $L\to\infty$, 
 then the Kraus operators \eq{\hat{K}_l=\sum_{M=0}^\infty\sum_{m=0}^M \ket{M,0}\bra{m,M-m}\frac{\eu^{\iu \tfrac{2\pi l m}{L}}}{\sqrt{L}}\label{eq:Kraus depolarize full}} form a CPTP map corresponding to the Mueller matrix
\eq{M=\begin{pmatrix}
		1&0&0&0\\
		0&0&0&0\\
		0&0&0&0\\
		1&0&0&0\\
\end{pmatrix}.} 

The insight gained from quantum mechanics is that the Kraus operators in \eqref{eq:Kraus depolarize full} are not physical. The true limit required is indeed $L\to\infty$, but requires maintaining $L\in\mathds{Z}$. While this bears resemblance to a discrete Fourier transform, the latter requires a substitution of $L\to M$ in \eqref{eq:Kraus depolarize full}. In the limit of $L\to\infty$, all of the Kraus operators  vanish, obliterating the potential for achieving the desired CPTP map. We see that the trace-preservation condition of quantum mechanics negates the possibility of certain depolarization Mueller matrices being valid for \textit{all} input states, which is typically an assumption of polarimetry. However, in realistic scenarios with a limited photon number, we see that certain ``nonphysical'' Mueller matrices become viable! This can help explain the validity of experimentally found Mueller matrices that seem to contradict known physical constraints (see, e.g., \cite{Simonetal2010}); perhaps the Mueller matrices found would not have been identical for other input states.

\section{Discussion and conclusions}
The main feature of quantum polarimetry is that the \add{quantum }Stokes \textit{operators} must transform according to Mueller calculus, regardless of the quantum state in question, for the former to agree with the predictions of classical polarimetry. A few simple quantum operations provide the building blocks for quantum polarimetry, and the degrees of freedom of an arbitrary Mueller matrix represent the various ways in which these building blocks can be combined.

One implication of the quantum channels described here is that all of polarimetry can be described in a trace-preserving manner. Total probability can always be conserved while explaining both deterministic and nondeterministic Mueller matrices, albeit with some information leaking to the environment about the states and transformations in question. The quantum channels can then be further decomposed into quantum master equations for the time evolution of systems evolving under nondepolarizing and depolarizing channels\add{; and, conversely, it can immediately be seen whether a given quantum master equation corresponds to a classical Mueller-matrix transformation}.

Another important result is the distinction between depolarizing and nondepolarizing processes, which has garnered much attention \cite{GilBernabeu1985,Simon1987,Simon1990,AndersonBarakat1994,CloudePottier1995,Gil2000,Zakerietal2013}. By assessing whether a process can be described by an operation that conserves photon number in an enlarged Hilbert space, one can immediately discern whether the associated Mueller matrix is nondepolarizing. All processes that require descriptions whereby modes in an enlarged Hilbert space are excited without de-exciting modes in the Hilbert space of $L$ and $R$ must be deemed depolarizing. We hope that this distinction proves useful in future studies of depolarization.

It is well-known that nondepolarizing Mueller matrices can be represented in the proper orthochronous Lorentz group \cite{Barakat1981}. However, there is no such constraint on the corresponding Kraus operators; the projections \eq{\hat{K}_l=\bra{l}_v\hat{U}_{LRv}\ket{0}_v =\bra{0}_v\hat{U}_{LRv}\ket{0}_v\frac{\left(\alpha_1\haL+\alpha_2\haR\right)^l}{\sqrt{l!}}} need not form a group ($\alpha_1$ and $\alpha_2$ are components of $\mathbf{U}_{LRv}$ as per Appendix \ref{app:explicit results}). For a $4\times 4$ Lorentz matrix $\Lambda$, we can find many decompositions \eq{\sum_l\hat{K}_l^\dagger\hat{S}_\mu \hat{K}_l=\sum_{\nu}\Lambda_{\mu\nu}\hat{S}_\nu,} and the Kraus operators conspire to yield a Lorentz transformation. 
We have shown that the classical degrees of freedom found in $\Lambda_{\mu\nu}$ correspond to the degrees of freedom in truncated unitary matrices acting on vectors of creation operators. Since this is equivalent to a $2\times 2$ complex matrix acting on $\left(\haLd,\haRd\right)^T$, it is equivalent to a $2\times 2$ complex Jones matrix $\mathbf{J}$ acting on $\mathbf{E}$ with the same matrix elements (up to a change of basis from linear to circular coordinate systems), and Jones matrices can indeed represent the Lorentz group. One arena in which this equivalence can be exploited is in the study of the aggregate properties of random nondepolarizing Mueller matrices.

We have conjectured above that quantum channels are responsible for all Mueller matrices being expressible as positive sums of nondepolarizing Mueller matrices \add{(Ref. \cite{Simonetal2010} has recently shown that only such sums are permissible for transformations of \textit{spatially-dependent} Stokes operators)}. A subset of the conjecture is that all quantum channels that take $\hat{S}_0\to\hat{S}_0$ while taking the other Stokes operators to linear combinations of Stokes operators must be able to be expressed as positive sums of nondepolarizing channels. This is related to studies of SU(2)-covariant channels, which give conditions for quantum channels to \textit{commute} with SU(2) operations \cite{KeylWerner1999,Boileauetal2008,GourSpekkens2008,Marvian2012,Streltsovetal2017}; specifically, this is related to channels for which \eq{\eu^{\iu \theta \hat{S}_0}\mathcal{E}\left(\rho\right)\eu^{-\iu \theta \hat{S}_0}=\mathcal{E}\left(\eu^{\iu \theta \hat{S}_0}\rho\eu^{-\iu \theta \hat{S}_0}\right),} which have interesting physical implications \cite{Morrisetal2019}. However, such channels imply not only $\hat{S}_0\to\hat{S}_0$ but also $\hat{S}_0^n\to\hat{S}_0^n$ for arbitrary integer $n$, which is a more stringent requirement than that of polarimetry. Inspection of the transformation \eq{\frac{\id_N}{N+1}\to \frac{1}{2}\left(\frac{\id_{N-1}}{N}+\frac{\id_{N+1}}{N+2}\right),} for example, shows the preservation of expectation values of $\hat{S}_0$ but not $\hat{S}_0^2$. This requirement automatically prevents the intermixing of subspaces with different photon numbers, whereas we conjecture that the intermixing of subspaces is prevented automatically by only limiting the linear transformations of Stokes operators (we expect this to be intimately linked with no-cloning arguments \cite{KeylWerner1999}). The connection to covariant channels may be an optimal starting point for proving our conjecture.

Polarimetry is concerned with \textit{measurement}; the goal is to characterize the changes effected by an intervening medium on an arbitrary input state. Armed with a quantum description of polarimetry, we can now discuss quantum strategies for estimating the elements of a Mueller matrix. It is well-known that quantum parameter estimation offers dramatic enhancements over classical parameter estimation \textit{for particular input states} (see \cite{SidhuKok2019}\add{\cite{Liuetal2019}} for \del{a} thorough recent review\add{s}). A necessary ingredient for finding such an enhancement is knowing how a transformed quantum state varies with the parameter being estimated.
This is an important application of our work; by explicitly finding an expression \eq{\hat{\rho}^\prime\left(\boldsymbol{\lambda}\right)=\mathcal{E}_{\boldsymbol{\lambda}}\left(\hat{\rho}\right)=\sum_l \hat{K}_l\left(\boldsymbol{\lambda}\right)\hat{\rho} \hat{K}_l^\dagger\left(\boldsymbol{\lambda}\right)} for the dependence of a transformed state $\hat{\rho}^\prime$ on a set of parameters $\boldsymbol{\lambda}$, we can investigate the states that are most sensitive to changes in the parameters being measured \add{\cite{Jietal2008}}. In the case of polarimetry, the parameters $\boldsymbol{\lambda}$ correspond to the 16 degrees of freedom of a Mueller matrix, and we can now ascertain how arbitrary quantum states vary with changes in Mueller matrix parameters. \add{This can be used in the quantum Fisher information paradigm to search for states with the \textit{potential} to be more sensitive to Mueller matrix parameters than classical polarization states without needing to specify a specific measurement scheme. For example, the usefulness of polarization-squeezed states in reducing fluctuations in parameter estimation can be investigated \textit{in vacuo} \cite{Alodzhantsetal1995,Alodjantsetal1998,Korolkovaetal2002}, similar to the usefulness of quadrature-squeezed light in interferometry \cite{Caves1981,Yurkeetal1986,Dowling1998}.}

Some relevant works have discussed quantum methods for simultaneously measuring a few degrees of freedom of a Mueller matrix. \add{Estimating a single rotation parameter is akin to phase estimation, which has been considerably optimized \cite{LIGO2011}, and estimating a single attenuation parameter can be equivalent to transmission measurements, for which all Fock states outperform classical states of light \cite{Adessoetal2009,Yoonetal2020}. } One study showed the tradeoff in measuring a single rotation angle around a known axis together with a single \del{depolarization }\add{attenuation } parameter \cite{Crowleyetal2014}, which can be cast into the product of Mueller matrices $M_R$ describing rotations around a known axis and $M_D$ describing \add{isotropic diattenuation [\eqref{eq:Diattenuator matrix example} with $q=r$]} \del{partial (isotropic) depolarization}. There, suitably-chosen quantum states offer enhanced sensitivities over classical measurements. More recently, quantum advantages have been for simultaneously measuring all three rotation parameters \cite{GoldbergJames2018Euler}. The search for quantum states and quantum estimation strategies that offer such advantages in the simultaneous estimation of all 16 degrees of freedom of a Mueller matrix certainly merits future study. 

Quantum channels allow for more sophisticated transformations than those described by Mueller matrices. This opens the door to new avenues of characterizing substances thought to be described by Mueller matrices, \del{though }\add{through } an analysis of the \textit{nonlinear} transformations of Stokes parameters enacted by a substance. The quantum polarimetry described here is a crucial building block for such analyses.

We have exhaustively shown how to describe classical polarimetry by transformations on quantum states. The components of Mueller matrices can now be directly related to the parameters of quantum channels, from which we ascertain the variety of quantum transformations of light that lead to the same classical measurement results. This underscores the importance of quantum polarization for improving classical measurements and better understanding the true nature of light.

\begin{acknowledgments}
	AZG acknowledges crucial conversations with Daniel James and Luis S\'anchez-Soto. This work was supported by the NSERC 
	Alexander Graham Bell Scholarship \#504825, Walter C. Sumner Foundation, and Lachlan Gilchrist Fellowship Fund.
\end{acknowledgments}

\appendix
\section{Explicit results}
\label{app:explicit results}
For the pure diattenuation described by \eqref{eq:diattenuation unitary 1} and Fig. \ref{fig:depolarization two vac} in the main text, one can write the corresponding transformation matrix $\mathbf{T}$:
\eq{\hat{U}_D^\dagger\begin{pmatrix}
		\hat{v}^\dagger_1\\\haLd\\\haRd\\\hat{v}_2^\dagger
\end{pmatrix}\hat{U}_D=\mathbf{T}\begin{pmatrix}
\hat{v}^\dagger_1\\\haLd\\\haRd\\\hat{v}_2^\dagger
\end{pmatrix}.} This $\mathbf{T}$ matrix is given by:
	\begin{widetext}\eq{\left(\begin{array}{cccc}
			\sqrt{q} & \eu^{-\frac{\iu\psi}{2}} \sqrt{1-q} \cos \left(\frac{\theta }{2}\right) & \eu^{\frac{\iu \psi }{2}} \sqrt{1-q} \sin \left(\frac{\theta }{2}\right) & 0 \\
			-\eu^{\frac{\iu \psi }{2}} \sqrt{1-q} \cos \left(\frac{\theta }{2}\right) & \frac{1}{2} \left[\sqrt{q}+\sqrt{r}+\left(\sqrt{q}-\sqrt{r}\right) \cos \left(\theta \right)\right] & \frac{1}{2} \eu^{\iu \psi } \left(\sqrt{q}-\sqrt{r}\right) \sin \left(\theta \right) & -\eu^{\frac{\iu \psi }{2}} \sqrt{1-r} \sin \left(\frac{\theta }{2}\right) \\
			-\eu^{-\frac{\iu\psi}{2}} \sqrt{1-q} \sin \left(\frac{\theta }{2}\right) & \frac{1}{2} \eu^{-\iu \psi } \left(\sqrt{q}-\sqrt{r}\right) \sin \left(\theta \right) & \frac{1}{2} \left[\sqrt{q}+\sqrt{r}+\left(\sqrt{r}-\sqrt{q}\right) \cos \left(\theta \right)\right] & \eu^{-\frac{\iu\psi}{2}} \sqrt{1-r} \cos \left(\frac{\theta }{2}\right) \\
			0 & \eu^{-\frac{\iu\psi}{2}} \sqrt{1-r} \sin \left(\frac{\theta }{2}\right) & -\eu^{\frac{\iu \psi }{2}} \sqrt{1-r} \cos \left(\frac{\theta }{2}\right) & \sqrt{r} \\
		\end{array}
		\right).}
	While $\mathbf{T}$ is unitary, its action on $\haLd$ and $\haRd$ after tracing out the two vacuum modes is not unitary, with the matrix in \eqref{eq:diattenuation on ab} corresponding only to the middle $2\times 2$ block of $\mathbf{T}$. The extra available degrees of freedom of the outer elements of $\mathbf{T}$ do not affect the polarimetric results. The Mueller matrix for this transformation is given by
	\eq{M_D=&\left(
		\begin{array}{cccc}
			\frac{q+r}{2} & \frac{1}{2} (q-r) \cos (\psi ) \sin (\theta )  \\
			\frac{1}{2} (q-r) \cos (\psi ) \sin (\theta ) & \frac{1}{4} \left\{q+r+2 \sqrt{q r}+\left[\cos ^2(\theta )-\cos (2 \psi ) \sin ^2(\theta )\right] \left(-q-r+2 \sqrt{q r}\right)\right\}  \\
			\frac{1}{2} (q-r) \sin (\theta ) \sin (\psi ) & \frac{1}{4} \left(q+r-2 \sqrt{q r}\right) \sin ^2(\theta ) \sin (2 \psi )  \\
			\frac{1}{2} (q-r) \cos (\theta ) & \frac{1}{4} \left(q+r-2 \sqrt{q r}\right) \cos (\psi ) \sin (2 \theta )  \\
		\end{array}
		\right.\\
	&\left.
	\begin{array}{cccc}
		\frac{1}{2} (q-r) \sin (\theta ) \sin (\psi ) & \frac{1}{2} (q-r) \cos (\theta ) \\
		 \frac{1}{4} \left(q+r-2 \sqrt{q r}\right) \sin ^2(\theta ) \sin (2 \psi ) & \frac{1}{4} \left(q+r-2 \sqrt{q r}\right) \cos (\psi ) \sin (2 \theta ) \\
		 \frac{1}{4} \left\{q+r+2 \sqrt{q r}+\left[\cos ^2(\theta )+\cos (2 \psi ) \sin ^2(\theta )\right] \left(-q-r+2 \sqrt{q r}\right)\right\} & \frac{1}{4} \left(q+r-2 \sqrt{q r}\right) \sin (2 \theta ) \sin (\psi ) \\
		 \frac{1}{4} \left(q+r-2 \sqrt{q r}\right) \sin (2 \theta ) \sin (\psi ) & \frac{1}{4} \left[q+r+2 \sqrt{q r}+\cos (2 \theta ) \left(q+r-2 \sqrt{q r}\right)\right] \\
	\end{array}
	\right).
}

For the SU(3) matrix $\mathbf{U}_{LRv}$ in \eqref{eq:diattenuator and retarder SU3} depicted schematically in Fig. \ref{fig:depolarization one vac} of the main text, we put the eight real degrees of freedom in the components $\alpha_1$, $\alpha_2$, $\beta_1$, and $\beta_2$:
\eq{\mathbf{U}_{LRv}=\begin{pmatrix}
		\alpha_1&\alpha_2&\sqrt{1-\left|\alpha_1\right|^2-\left|\alpha_2\right|^2}\eu^{\iu \theta_{a3}}\\
		\beta_1&\beta_2&\sqrt{1-\left|\beta_1\right|^2-\left|\beta_2\right|^2}\eu^{\iu \theta_{b3}}\\
		\sqrt{1-\left|\alpha_1\right|^2-\left|\beta_1\right|^2}\eu^{\iu \theta_{v1}}&\sqrt{1-\left|\alpha_2\right|^2-\left|\beta_2\right|^2}\eu^{\iu \theta_{v2}}&\sqrt{\left|\alpha_1\right|^2+\left|\beta_1\right|^2+\left|\alpha_2\right|^2+\left|\beta_2\right|^2-1}\eu^{\iu \theta_{v3}}\\
\end{pmatrix}.} The five angles are determined by the orthogonality equations between each pair of rows and columns, of which one is redundant; for example,
\eq{\alpha_1\beta_1^*+\alpha_2\beta_2^*+\sqrt{\left(1-\left|\alpha_1\right|^2-\left|\alpha_2\right|^2\right)\left(1-\left|\beta_1\right|^2-\left|\beta_2\right|^2\right)}\eu^{\iu\left(\theta_{a3}-\theta_{b3}\right)}=0.} There are no remaining degrees of freedom in this minimal description of a quantum channel corresponding to a nondepolarizing Mueller matrix.
	
\end{widetext}

\end{document}